\documentclass[preprint,superscriptaddress,nofootinbib]{revtex4}
  
  \usepackage{amssymb,amsmath,graphicx,subfigure,color}
  \usepackage[colorlinks,
              linkcolor=red, 
              citecolor=cyan
              ]{hyperref}
  \allowdisplaybreaks[4]
  \begin{document}

  \title{Feasibility of searching for the Cabibbo-favored \\
         $D^{\ast}$ ${\to}$ $\overline{K}{\pi}^{+}$,
         $\overline{K}^{\ast}{\pi}^{+}$,
         $\overline{K}{\rho}^{+}$ decays}
  \author{Yueling Yang}
  \affiliation{Institute of Particle and Nuclear Physics,
              Henan Normal University, Xinxiang 453007, China}
  \author{Kang Li}
  \affiliation{Institute of Particle and Nuclear Physics,
              Henan Normal University, Xinxiang 453007, China}
  \author{Zhenglin Li}
  \affiliation{Institute of Particle and Nuclear Physics,
              Henan Normal University, Xinxiang 453007, China}
  \author{Jinshu Huang}
  \affiliation{School of Physics and Electronic Engineering,
              Nanyang Normal University, Nanyang 473061, China}
  \author{Qin Chang}
  \affiliation{Institute of Particle and Nuclear Physics,
              Henan Normal University, Xinxiang 453007, China}
  \author{Junfeng Sun}
  \affiliation{Institute of Particle and Nuclear Physics,
              Henan Normal University, Xinxiang 453007, China}

   \begin{abstract}
   The current knowledge on the $D^{\ast}$ mesons are still
   inadequate.
   Encouraged by the positive development prospects of
   high-luminosity and high-precision experiments,
   the Cabibbo-favored nonleptonic $D^{\ast}$ ${\to}$
   $\overline{K}{\pi}^{+}$,  $\overline{K}^{\ast}{\pi}^{+}$,
   $\overline{K}{\rho}^{+}$ weak decays are studied
   with the naive factorization approach.
   It is found that branching ratios of these processes
   can reach up to ${\cal O}(10^{-10})$ or more,
   and can be accessible at STCF, CEPC, FCC-ee and
   LHCb@HL-LHC experiments in the future.
   It might even be possible to search for the
   $D^{{\ast}0}$ ${\to}$ $K^{{\ast}-}{\pi}^{+}$ and
   $K^{-}{\rho}^{+}$ decays
   at the running SuperKEKB experiments.

   \href{https://link.aps.org/doi/10.1103/PhysRevD.106.036029}
        {Phys. Rev. D 106, 036029 (2022)}
   \end{abstract}
   \keywords{$D^{\ast}$ meson; weak decay; branching ratio.}

  \maketitle

   \section{Introduction}
   \label{sec01}
\begin{figure}[h]
   \includegraphics[width=0.95\textwidth]{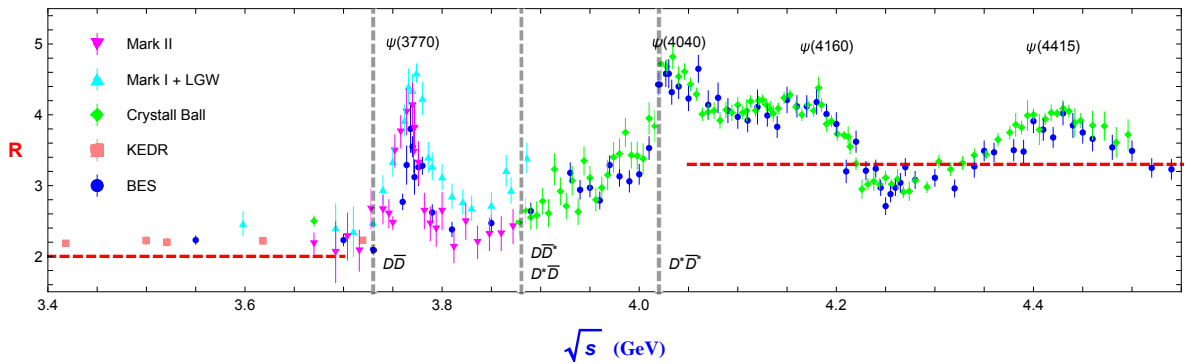}
   \vspace{-6mm}
   \caption{$R$ near the charm threshold regions versus
   the center-of-mass energy $\sqrt{s}$ of $e^{+}e^{-}$
   annihilation, corresponding to Fig.53.3 of Ref. \cite{pdg2020},
   where the experimental data are available at web page
   https://pdg.lbl.gov/2021/hadronic-xsections  \cite{pdg2020},
   the lower and upper horizontal dashed lines correspond to
   $R$ $=$ $2$ and $10/3$, respectively; and the vertical
   dashed lines denote charm thresholds.}
   \label{fig-r}
   \end{figure}

   The need for a charm quark as its natural incorporation in
   a unified description of the weak and electromagnetic
   interactions, was first postulated by Bjorken, Glashow,
   Iliopoulos, and Maiani \cite{Phys.Lett.11.255,PhysRevD.2.1285}.
   The charm flavor is strictly conserved in
   electromagnetic and strong interactions.
   Although the $J/{\psi}$ particle was first discovered by
   two vastly different experiments in 1974
   \cite{Phys.Rev.Lett.33.1404,Phys.Rev.Lett.33.1406},
   and interpreted as a bound system of the $c\bar{c}$ pair,
   the charm flavor is hidden away in the quantity $R$ near
   the narrow resonance $J/{\psi}$ peak.
   The observable quantity $R$ measures the total hadron production
   rate relative to the electrodynamic ${\mu}^{+}{\mu}^{-}$
   process in $e^{+}e^{-}$ collisions, $R$ ${\equiv}$
   ${\sigma}(e^{+}e^{-}{\to}{\rm hadrons})
   /{\sigma}(e^{+}e^{-}{\to}{\mu}^{+}{\mu}^{-})$,
   determines the properties of the species of quark carrying
   color and fractional electric charge.
   The value of $R$ with the lowest order approximation should
   remain approximately constant with center-of-mass energy of
   the $e^{+}e^{-}$ pair annihilation, as long as no new final
   state fermion pair appears.
   From the view of $R$, the most striking evidence for
   the existence of a charm quark is the discovery of charmed
   mesons in positron-electron annihilation at SPEAR experiments
   in 1976 \cite{Phys.Rev.Lett.37.255,Phys.Rev.Lett.37.569}.
   The clearly evident increase in $R$ near the $4$ GeV region
   in Fig. \ref{fig-r} has been widely taken by particle
   physicists to be the characteristic features of charm threshold.

\begin{figure}[h]
   \includegraphics[width=0.4\textwidth]{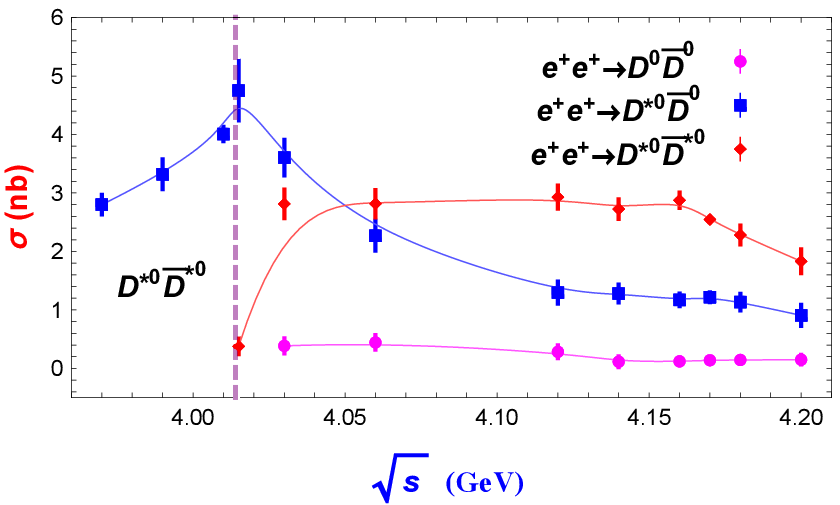} \quad
   \includegraphics[width=0.4\textwidth]{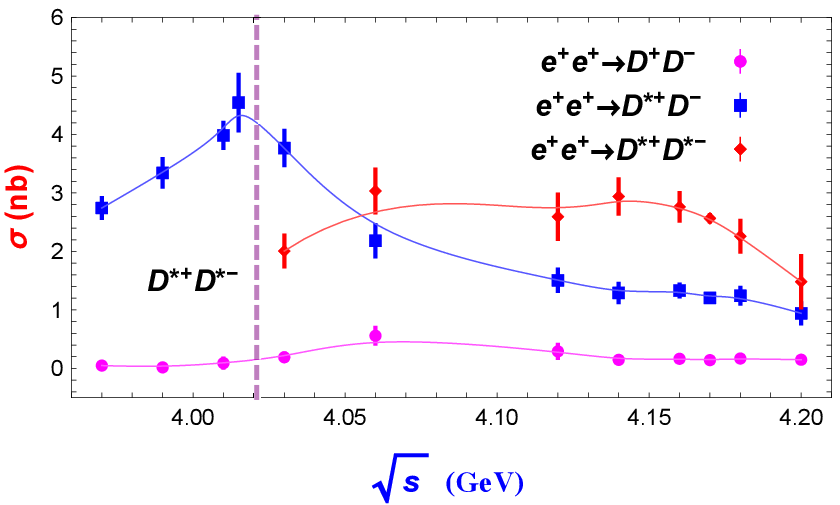}
   \vspace{-6mm}
   \caption{Measured production cross sections for charmed meson
   pairs
   near the threshold regions versus the center-of-mass
   energy $\sqrt{s}$ of $e^{+}e^{-}$ annihilation,
   where the experimental data are available from the CLEO-c
   group in Refs. \cite{PhysRevD.80.072001,cpc.42.043002};
   the vertical dashed lines denote charm thresholds.}
   \label{fig-cs}
   \end{figure}

   The ordinary charmed mesons consist of the charmed quark plus
   a light nonstrange quark forming an isodoublet, $c\bar{u}$
   and $c\bar{d}$, symbolized respectively by $D^{0}$ and $D^{+}$
   corresponding to the $1^{1}S_{0}$ state with spin-parity
   quantum $J^{P}$ $=$ $0^{-}$ for ground pseudoscalar mesons,
   $D^{{\ast}0}$ and $D^{{\ast}+}$  corresponding
   to the $1^{3}S_{1}$ state with spin-parity quantum $J^{P}$
   $=$ $1^{-}$ for ground vector mesons.
   In electron-positron collisions, the charmed quark pair
   originated from the virtual timelike photon combines
   with the light $u$ and $d$ quarks out of the vacuum with
   the orbit-spin $L$-$S$ coupling and forms the charmed
   mesons through strong interactions.
   In principle, the charmed meson pairs are equally likely
   to be neutral and charged.
   It is clearly seen from Fig. \ref{fig-r} that there are many
   broader resonances above charm threshold, such as
   ${\psi}(3770)$, ${\psi}(4040)$, ${\psi}(4160)$ and so on.
   These broader sibling ${\psi}$ resonances presumably decay
   strongly to pairs of oppositely charmed mesons,
   $D\bar{D}$, $D\bar{D}^{\ast}$,
   and $D^{\ast}\bar{D}^{\ast}$.
   The relative production cross sections of charmed mesons
   close to threshold were once thought to be, for example,
   ${\sigma}(D\bar{D})$:${\sigma}(D\bar{D}^{\ast})$:${\sigma}(D^{\ast}\bar{D}^{\ast})$
   ${\simeq}$ 1:4:7 \cite{PhysRevLett.37.477,PhysRevLett.37.398,
    Phys.Lett.B.65.55,Phys.Lett.B.66.70} or other ratios
    \cite{Phys.Lett.B.70.106,PhysRevD.15.759,PhysRevD.16.1327,
    PhysRevD.17.765}
   based on different theoretical calculation.
   The measured production ratio may differ considerably
   from those theoretical estimates.
   At energies a little far above the $D^{\ast}\bar{D}^{\ast}$
   threshold, it is generally believed that the production
   cross sections of $D^{\ast}$ mesons are greater than
   those of $D$ mesons in $e^{+}e^{-}$ collisions,
   which is clearly seen from Fig. \ref{fig-cs}.

   The center-of-mass energy for $D^{\ast}\bar{D}^{\ast}$
   production is larger than that for $D\bar{D}$ production.
   Due to the engineering design of $e^{+}e^{-}$
   accelerators and detectors, plus the actual operation
   energy and time, more experimental data on the
   pseudoscalar $D$ mesons rather than the vector
   $D^{\ast}$ mesons have been accumulated at CLEC-c
   and BES experiments.
   The properties of $D$ mesons have been extensively
   and carefully studied by many experimental groups.
   To go along with this, the experimental information on
   $D^{\ast}$ mesons is still very limited by now \cite{pdg2020}.
   The measurement of the mass of $D^{\ast}$ mesons which
   was quoted but not adopted by Particle Data Group was
   carried out in 1977 \cite{PhysRevLett.39.1301,Phys.Lett.B.69.503},
   not updated for 45 years.
   In addition, only three decay modes of $D^{{\ast}0}$
   ($D^{{\ast}+}$) mesons have been observed \cite{pdg2020}.

   The fitted $D^{\ast}$ masses are
   $m_{D^{{\ast}0}}$ $=$ $2006.85(5)$ MeV and
   $m_{D^{{\ast}+}}$ $=$ $2010.26(5)$ MeV \cite{pdg2020}.
   There is a hierarchical relationship among hadron mass,
     \begin{equation}
     m_{D^{{\ast}0}} - m_{D^{0}}\,  > \, m_{{\pi}^{0}}
     \label{mass-splitting-cu-cu},
     \end{equation}
     \begin{equation}
     m_{D^{{\ast}0}} - m_{D^{+}}\, < \, m_{{\pi}^{-}}
     \label{mass-splitting-cu-cd},
     \end{equation}
     \begin{equation}
     m_{D^{{\ast}+}} - m_{D^{0}}\, > \, m_{{\pi}^{+}}
     \label{mass-splitting-cd-cu},
     \end{equation}
     \begin{equation}
     m_{D^{{\ast}+}} - m_{D^{+}}\, > \, m_{{\pi}^{0}}
     \label{mass-splitting-cd-cd},
     \end{equation}
   which results in the $D^{{\ast}0}$ ${\to}$ $D^{+}{\pi}^{-}$
   decay being kinematically forbidden.
   The $D^{{\ast}}$ mesons can decay mainly through the strong
   and electromagnetic interactions, and are dominated
   by $D^{\ast}$ ${\to}$ $D{\pi}$ processes.
   Kinematically, the $D^{\ast}$ ${\to}$ $D{\pi}$ decays
   are inhibited by very compact phase space.
   Dynamically, for the $D^{\ast}$ ${\to}$ $D{\pi}$ decays,
   the ${\pi}$ emission processes are suppressed by the
   phenomenological Okubo–Zweig–Iizuka rule
   \cite{ozi-o,ozi-z,ozi-i},
   and the recoiled ${\pi}$ processes are suppressed by a
   color match factor owing to the color singlet requirements
   of quark combinations.
   Consequently, the decay width of $D^{\ast}$ mesons is
   very narrow, for example, ${\Gamma}_{D^{{\ast}+}}$ $=$
   $83.4{\pm}1.8$ keV \cite{pdg2020}.
   Experimentally, the momentum of the pion in the rest frame
   of $D^{\ast}$ mesons is very soft, about $40$ MeV.
   The difficulties in signal reconstruction result in
   inefficiency of particle identification.

   Besides, the $D^{{\ast}}$ meson decay through weak interactions
   is also legal and allowable within the standard model (SM) of
   elementary particle.
   As it is well known, the strong and electromagnetic
   interactions are related only to vector currents,
   while the weak interactions are related to both
   vector and axial vector currents.
   Study of the $D^{{\ast}}$ meson weak decays can not only
   improve our knowledge of the properties $D^{{\ast}}$ mesons,
   but also test the axial vector current interactions in SM
   and enrich our understanding on the decay mechanism of
   $D^{{\ast}}$ mesons.
   In addition, the Cabibbo-Kobayashi-Maskawa (CKM) matrix elements
   \cite{PhysRevLett.10.531,PTP.49.652} describing the quark mixing
   can be determined and overconstrained from $D^{{\ast}}$
   meson weak decays.

   It is not hard to visualize that the occurrence probability
   of $D^{{\ast}}$ meson weak decays is very small,
   insignificant when compared with that of strong and
   electromagnetic decays.
   Fortunately, the experimental data on $D^{{\ast}}$ mesons
   are accumulating increasingly.
   Now, more than $5{\times}10^{7}$ $D^{{\ast}{\pm}}$ mesons
   have been accumulated in energy regions $\sqrt{s}$ ${\in}$
   $[4.1,4.6]$ at BESIII experiments \cite{JHEP2022.05.155}.
   It is promisingly expected that there will be a total of
   about $5{\times}10^{10}$ $c\bar{c}$ pairs at the
   SuperKEKB \cite{PTEP.2019.123C01}.
   Given the charm quark fragmentation fractions
   $f(c{\to}D^{{\ast}+})$ ${\approx}$ $25\%$ and
   $f(c{\to}D^{{\ast}0})$ ${\approx}$ $23\%$ \cite{epjc.76.397},
   more than $2{\times}10^{10}$ $D^{{\ast}0}$ and
   $D^{{\ast}{\pm}}$
   will be available at SuperKEKB.
   If it is considered optimistically that the production cross sections
   ${\sigma}(e^{+}e^{-}{\to}D^{0}\bar{D}^{{\ast}0})$ ${\approx}$
   ${\sigma}(e^{+}e^{-}{\to}D^{+}D^{{\ast}-})$ ${\approx}$
   $2$ nb and
   ${\sigma}(e^{+}e^{-}{\to}D^{{\ast}0}\bar{D}^{{\ast}0})$ ${\approx}$
   ${\sigma}(e^{+}e^{-}{\to}D^{{\ast}+}D^{{\ast}-})$ ${\approx}$
   $3$ nb near ${\psi}(4040)$ resonance,
   as illustrated in Fig. \ref{fig-cs} (or Table 1 and 2
   of Ref. \cite{JHEP2022.05.155}),
   about $8{\times}10^{10}$ $D^{{\ast}0}$ and
   $D^{{\ast}{\pm}}$ mesons will be available
   at the super ${\tau}$-charm factory (STCF)
   \cite{STCF,SCTF} with a total integrated luminosity
   of $10\,{\rm ab}^{-1}$.
   It is highly expected that there will be a total of
   about $10^{12}$ $Z^{0}$ bosons at the Circular Electron
   Positron Collider (CEPC) \cite{cepc} and
   $10^{13}$ $Z^{0}$ bosons at the Future Circular
   Collider (FCC-ee) \cite{fcc}.
   Considering the inclusive branching ratios
   ${\cal B}(Z^{0}{\to}D^{{\ast}{\pm}}X)$ $=$
   $(11.4{\pm}1.3)\%$ \cite{pdg2020} and the approximation
   ${\cal B}(Z^{0}{\to}D^{{\ast}0}X/\bar{D}^{{\ast}0}X)$
   ${\approx}$ ${\cal B}(Z^{0}{\to}D^{{\ast}{\pm}}X)$,
   more than $10^{11}$ and $10^{12}$ $D^{{\ast}}$ mesons
   will be available at CEPC and FCC-ee with a total
   integrated luminosity of $20\,{\rm ab}^{-1}$,
   respectively.
   In addition, the inclusive production cross sections in
   hadron-hadron collisions are measured to be
   ${\sigma}(pp{\to}D^{{\ast}+}X)$ ${\approx}$ $0.8$ mb
   at center-of-mass energy of $\sqrt{s}$ $=$ $13$ TeV
   by the LHCb group \cite{JHEP.2016.03.159}
   and ${\sigma}_{\rm tot}(D^{{\ast}+})$ ${\approx}$
   $2.1$ mb at $\sqrt{s}$ $=$ $7$ TeV by the ALICE group
   \cite{JHEP.2012.07.191}, respectively.
   A conservative estimate is that more than
   $2{\times}10^{14}$ $D^{{\ast}}$ mesons will be available
   by the LHCb detector at the future High Luminosity
   LHC (HL-LHC) hadron collider with a total integrated
   luminosity of $300\,{\rm fb}^{-1}$ \cite{1808.08865}.
   A huge amount of experimental data provides a solid
   foundation and valuable opportunities for studying
   and understanding the properties of $D^{{\ast}}$
   mesons, including the search for $D^{{\ast}}$ meson
   weak decays.

   It is natural to want to know the probability
   of $D^{{\ast}}$ meson weak decays, and whether the study
   on the $D^{{\ast}}$ meson weak decays is feasible or not
   in the coming future.
   Our study \cite{Eur.Phys.J.C.81.1110}
   has tentatively shown that the purely leptonic
   $D^{{\ast}+}$ decays might be measurable at $e^{+}e^{-}$
   colliders, such as SuperKEKB, STCF, CEPC and FCC-ee,
   and HL-LHC hadron collider \cite{1808.08865}.
   Purely leptonic $D^{{\ast}+}$ decays are Cabibbo-suppressed,
   while leptonic $D^{{\ast}0}$ decays are induced by the
   flavor-changing neutral weak currents.
   In this paper, we would like to investigate the
   $D^{{\ast}}$ ${\to}$ $\overline{K}{\pi}^{+}$,
   $\overline{K}{\rho}^{+}$ and
   $\overline{K}^{\ast}{\pi}^{+}$ decays within the SM.
   The motivation is as follows.
   Dynamically, these decays are induced by $W$ boson
   emission, and favored by CKM elements, and thus should have
   relatively larger branching ratios and higher priority to
   do research among $D^{{\ast}}$ meson weak decays.
   Kinematically, the final states are back-to-back,
   and have definitive momentum and discrete energy in the
   rest frame of $D^{{\ast}}$ mesons.
   Experimentally, the energetic charged pion and kaon of final
   states of nonleptonic $D^{{\ast}}$ weak decays are more
   easily captured by the sensitive detectors, when compared
   with $D^{\ast}$ ${\to}$ $D{\pi}$ decays where the soft pion
   identifications are very challenging, and when compared with
   leptonic $D^{{\ast}+}$ decays where the final neutrinos
   will bring the signal reconstruction into additional
   complications.
   Hence, there would be a relatively higher reconstruction
   efficiency for nonleptonic $D^{{\ast}}$ weak decays.

   The remainder of this paper is organized as follows.
   Theoretical framework and phenomenological approach
   dealing with nonleptonic $D^{{\ast}}$ weak decays are
   presented in Section \ref{sec:hamiltonian}.
   Numerical results of branching ratios and our comments
   are given in Section \ref{sec:branch}.
   Section \ref{sec:summary} is a brief summary.
   The analytic expressions of decay amplitudes are
   collected in Appendixes \ref{sec:v2pp} and \ref{sec:v2vp}.

   \section{Theoretical framework}
   \label{sec:hamiltonian}
   The effective Hamiltonian in charge of nonleptonic
   $D^{{\ast}}$ ${\to}$ $\overline{K}{\pi}^{+}$,
   $\overline{K}{\rho}^{+}$ and
   $\overline{K}^{\ast}{\pi}^{+}$ decays
   is written as \cite{RevModPhys.68.1125},
   \begin{equation}
  {\cal H}_{\rm eff}\, =\,
   \frac{G_{F}}{\sqrt{2}}\, V_{cs}^{\ast}\,V_{ud}\, \big\{
    C_{1}\,O_{1}+C_{2}\,O_{2} \big\} + {\rm h.c.}
   \label{hamilton},
   \end{equation}
   where both Fermi constant $G_{F}$ ${\approx}$
   $1.166{\times}10^{-5}$ ${\rm GeV}^{-2}$ \cite{pdg2020}
   and Wilson coefficients $C_{1,2}$ are
   universal effective couplings.
   Wilson coefficients are computable with the perturbation
   theory at the mass scale of $W$ boson.
   The scale-dependence of Wilson coefficients can be
   obtained with the renormalization group equation
   \cite{RevModPhys.68.1125}.
   The nature of the product of $G_{F}\,C_{i}$ is similar
   to a gauge parameter, for example, the electric
   charge $e$ for electromagnetic interactions.
   The CKM elements correspond to different effective
   operators, and have been determined precisely,
   ${\vert}V_{cs}{\vert}$ $=$ $0.987(11)$, and
   ${\vert}V_{ud}{\vert}$ $=$ $0.97370(14)$ \cite{pdg2020}.
   The explicit expressions of effective four-quark
   operators are written as,
     \begin{eqnarray}
     O_{1} &=&
     \big[ \bar{s}_{\alpha}\,{\gamma}^{\mu}\,(1-{\gamma}_{5})\,c_{\alpha} \big]\,
     \big[ \bar{u}_{\beta}\,{\gamma}_{\mu}\,(1-{\gamma}_{5})\,d_{\beta} \big]
     \label{operator-o1},  \\
     O_{2} &=&
     \big[ \bar{s}_{\alpha}\,{\gamma}^{\mu}\,(1-{\gamma}_{5})\,c_{\beta} \big]\,
     \big[ \bar{u}_{\beta}\,{\gamma}_{\mu}\,(1-{\gamma}_{5})\,d_{\alpha} \big]
     \label{operator-o2},
     \end{eqnarray}
    where ${\alpha}$ and ${\beta}$ are the color indices.
    Owing to only one kaon meson in final states, there
    is no penguin operators.

    Taking the $D^{{\ast}0}$ ${\to}$ $K^{-}{\pi}^{+}$
    decay for example, the amplitude can be written as,
     \begin{equation}
    {\cal A}(D^{{\ast}0}{\to}K^{-}{\pi}^{+})\, =\,
    {\langle}K^{-}{\pi}^{+}{\vert}
    {\cal H}_{\rm eff}
    {\vert} D^{{\ast}0} {\rangle}\, =\,
     \frac{G_{F}}{\sqrt{2}}\, V_{cs}^{\ast}\,V_{ud}\,
     \sum\limits_{i=1}^{2}C_{i}\,
    {\langle}K^{-}{\pi}^{+}{\vert} O_{i}
    {\vert} D^{{\ast}0} {\rangle}
     \label{Hamiltonian-amplitude}.
     \end{equation}
    Clearly, the remaining part of decay amplitudes is to
    reasonably evaluate the hadronic matrix elements (HMEs)
    ${\langle}K^{-}{\pi}^{+}{\vert}O_{i}
     {\vert} D^{{\ast}0} {\rangle}$, which sandwich the
    initial and final states together with quark operators.
    HMEs closely relate to transitions between quarks and
    hadron states, and involve short- and long-distance
    contributions which complicate theoretical calculations.

    Phenomenologically, the naive factorization (NF) approach
    \cite{zpc.34.103} was often used to deal with HMEs,
    owing to its simple and clear physics picture, and
    good performance for nonleptonic $B$ and $D$ weak decays
    induced by external $W$ emission processes.
    Based on the color transparency hypothesis \cite{npbps.11.325}
    that energetic final hadron states have flown far away from
    each other before influence of soft gluons, HMEs of four-quark
    operators using NF approach are divided into two HMEs of
    hadron currents, and the final state interactions can be
    dismissed for the time being.
    HMEs of hadron currents are conventionally parametrized
    by decay constants and hadron transition form factors,
    which can be obtained from data and nonperturbative
    methods.
    In addition, it has been shown in Ref. \cite{PhysRevD.81.074021}
    that although the contributions from
   $W$-exchange and $W$-annihilation diagrams
   are the same order as those from tree-emission
   diagrams due to the large non-perturbative
   contributions for charmed meson decays,
   the tree-emission diagrams always give the
   largest contributions in the total amplitudes,
   thereby give the correct orders of magnitude of
   branching fractions.
    In this paper, we will apply NF approach to the concerned
    nonleptonic $D^{\ast}$ weak decays.

    The parametrization schemes of hadron current HMEs are
    generally written as
    \cite{zpc.29.637,ijmpa.30.1550094,jhep.1912.102},
     \begin{eqnarray}
    {\langle}0{\vert} \bar{q}_{1}\,{\gamma}_{\mu}\,q_{2}
    {\vert}P(k){\rangle} &=& 0
     \label{decay-pseudoscalar-v},
     \\
    {\langle}0{\vert} \bar{q}_{1}\,{\gamma}_{\mu}\,
    {\gamma}_{5}\,q_{2}{\vert}P(k){\rangle} &=& i\,f_{P}\,k_{\mu}
     \label{decay-pseudoscalar-a},
     \\
    {\langle}0{\vert} \bar{q}_{1}\,{\gamma}_{\mu}\,q_{2}
    {\vert}V(k,{\epsilon}){\rangle} &=& f_{V}\,m_{V}\,{\epsilon}_{\mu}
     \label{decay-vector-v},
     \\
    {\langle}0{\vert} \bar{q}_{1}\,{\gamma}_{\mu}\,
    {\gamma}_{5}\,q_{2}{\vert}V(k,{\epsilon}){\rangle} &=& 0
     \label{decay-vector-a},
     \end{eqnarray}
     \begin{equation}
    {\langle}P(p_{2}){\vert}\,\bar{q}\,{\gamma}_{\mu}\,c\,
    {\vert}D^{\ast}({\epsilon}_{1},p_{1}){\rangle}\, =\,
    {\varepsilon}_{{\mu}{\nu}{\alpha}{\beta}}\,
    {\epsilon}_{1}^{\nu}\,  P^{\alpha}\, q^{\beta}
     \frac{ V^{D^{\ast}P}(q^{2}) }{ m_{D^{\ast}}+m_{P} }
    \label{fromfactor-v2p-v},
    \end{equation}
     \begin{eqnarray} & &
    {\langle}P(p_{2}){\vert}\, \bar{q}\,{\gamma}_{\mu}\,
    {\gamma}_{5}\,c\,{\vert}D^{\ast}({\epsilon}_{1},p_{1})
    {\rangle}
    \nonumber \\ &=&
   +i\,2\,m_{D^{\ast}}\,
     \frac{ {\epsilon}_{1}{\cdot}q }{q^{2}}\,
     q_{\mu}\, A_{0}^{D^{\ast}P}(q^{2})
   +i\,{\epsilon}_{1,{\mu}}\, ( m_{D^{\ast}}+m_{P} )\,
     A_{1}^{D^{\ast}P}(q^{2})
     \nonumber \\ & &
   +i\,\frac{ {\epsilon}_{1}{\cdot}q }{ m_{D^{\ast}}+m_{P} }\,
     P_{\mu}\, A_{2}^{D^{\ast}P}(q^{2})
   -i\,2\,m_{D^{\ast}}\,
     \frac{ {\epsilon}_{1}{\cdot}q }{q^{2}}\,
     q_{\mu}\, A_{3}^{D^{\ast}P}(q^{2})
    \label{fromfactor-v2p-a},
    \end{eqnarray}
     \begin{eqnarray} & &
    {\langle}V({\epsilon}_{2},p_{2}){\vert}\,
     \bar{q}\,{\gamma}_{\mu}\,c\,
    {\vert}D^{\ast}({\epsilon}_{1},p_{1}){\rangle}\, =\,
   -({\epsilon}_{1}{\cdot}{\epsilon}_{2}^{\ast})\,
    \big\{ P_{\mu}\,V_{1}^{D^{\ast}V}(q^{2})
          -q_{\mu}\,V_{2}^{D^{\ast}V}(q^{2}) \big\}
    \nonumber \\ & &
   +\frac{ ({\epsilon}_{1}{\cdot}q)\,
           ({\epsilon}_{2}^{\ast}{\cdot}q) }
         { m_{D^{\ast}}^{2}-m_{V}^{2} }\,
    \big\{ \big[ P_{\mu} -
    \frac{ m_{D^{\ast}}^{2}-m_{V}^{2} }{ q^{2} }\,q_{\mu}
    \big]\, V_{3}^{D^{\ast}V}(q^{2})
   +\frac{ m_{D^{\ast}}^{2}-m_{V}^{2} }{ q^{2} }\,
           q_{\mu}\, V_{4}^{D^{\ast}V}(q^{2}) \big\}
     \nonumber \\ & &
    -({\epsilon}_{1}{\cdot}q)\,
     {\epsilon}_{2,{\mu}}^{\ast}\,V_{5}^{D^{\ast}V}(q^{2})
   +({\epsilon}_{2}^{\ast}{\cdot}q)\,
     {\epsilon}_{1,{\mu}}\,V_{6}^{D^{\ast}V}(q^{2})
    \label{fromfactor-v2v-v},
    \end{eqnarray}
     \begin{eqnarray} & &
   {\langle}V({\epsilon}_{2},p_{2}){\vert}\,
    \bar{q}\,{\gamma}_{\mu}\,{\gamma}_{5}\,c\,
   {\vert}D^{\ast}({\epsilon}_{1},p_{1}){\rangle}
    \nonumber \\ &=&
   -i\,{\varepsilon}_{{\mu}{\nu}{\alpha}{\beta}}\,
   {\epsilon}_{1}^{\alpha}\,
   {\epsilon}_{2}^{{\ast}{\beta}}\,\big\{ \big[ P^{\nu} -
    \frac{ m_{D^{\ast}}^{2}-m_{V}^{2} }{ q^{2} }\,q^{\nu}
    \big]\, A_{1}^{D^{\ast}V}(q^{2})
   +\frac{ m_{D^{\ast}}^{2}-m_{V}^{2} }{ q^{2} }\,
    q^{\nu}\, A_{2}^{D^{\ast}V}(q^{2}) \big\}
    \nonumber \\ & &
   -\frac{ i\,{\varepsilon}_{{\mu}{\nu}{\alpha}{\beta}}\,
           P^{\alpha}\, q^{\beta} }
         { m_{D^{\ast}}^{2}-m_{V}^{2} }\, \big\{
    ({\epsilon}_{2}^{\ast}{\cdot}q)\,
   {\epsilon}_{1}^{\nu}\,A_{3}^{D^{\ast}V}(q^{2})
  - ({\epsilon}_{1}{\cdot}q)\,
   {\epsilon}_{2}^{{\ast},{\nu}}\,A_{4}^{D^{\ast}V}(q^{2}) \big\}
    \label{fromfactor-v2v-a},
    \end{eqnarray}
   where $m_{P}$ and $f_{P}$ are the mass and decay constant
   of pseudoscalar $P$ meson, respectively;
   $m_{V}$, $f_{V}$, and ${\epsilon}_{V}$ are the mass,
   decay constant, and polarization vector of vector $V$
   meson, respectively;
   $A_{i}^{D^{\ast}h}$ and $V_{i}^{D^{\ast}h}$ are mesonic
   transition form factors;
   The momentum $P$ $=$ $p_{1}$ $+$ $p_{2}$ and
   $q$ $=$ $p_{1}$ $-$ $p_{2}$.
   There are some relationships among form factors,
    \begin{equation}
     (m_{D^{\ast}}+m_{P})\, A_{1}^{D^{\ast}P}(q^{2})
    +(m_{D^{\ast}}-m_{P})\, A_{2}^{D^{\ast}P}(q^{2})
     \, =\,
    2\,m_{D^{\ast}}\,A_{3}^{D^{\ast}P}(q^{2})
    \label{v2p-a1-a2-a3},
    \end{equation}
    \begin{equation}
    A_{0}^{D^{\ast}P}(0)\, =\, A_{3}^{D^{\ast}P}(0)
    \label{v2p-a0-a3},
    \end{equation}
    \begin{equation}
    A_{1}^{D^{\ast}V}(0)\, =\, A_{2}^{D^{\ast}V}(0)
    \label{v2v-a0-a2},
    \end{equation}
    \begin{equation}
    V_{3}^{D^{\ast}V}(0)\, =\, V_{4}^{D^{\ast}V}(0)
    \label{v2v-v3-v4}.
    \end{equation}

   The numerical values of form factors have been comprehensively
   calculated with the light front quark model in Ref. \cite{jhep.1912.102},
   and are listed in Table \ref{tab:formfactor}.
  \begin{table}[th]
  \caption{Numerical values of decay constants
  \cite{pdg2020,Eur.Phys.J.C.81.1110}
  and form factors at the pole $q^{2}$ $=$ $0$
  \cite{jhep.1912.102}.}
  \label{tab:formfactor}
  \begin{ruledtabular}
  \begin{tabular}{cccc}
     $f_{\pi}$      $=$ $130.2{\pm}1.2$ MeV
   & $f_{K}$        $=$ $155.7{\pm}0.3$ MeV
   & $f_{\rho}$     $=$ $207.7{\pm}1.6$ MeV
   & $f_{K^{\ast}}$ $=$ $202.5^{+6.5}_{-6.7}$ MeV \\ \hline
     $V_{1}^{D^{\ast}{\rho}}$ $=$ $0.65$
   & $V_{2}^{D^{\ast}{\rho}}$ $=$ $0.51$
   & $V_{3}^{D^{\ast}{\rho}}$ $=$ $0.29$
   & $V_{5}^{D^{\ast}{\rho}}$ $=$ $1.42$ \\
     $V_{6}^{D^{\ast}{\rho}}$ $=$ $0.68$
   & $A_{1}^{D^{\ast}{\rho}}$ $=$ $0.59$
   & $A_{3}^{D^{\ast}{\rho}}$ $=$ $0.22$
   & $A_{4}^{D^{\ast}{\rho}}$ $=$ $0.24$ \\ \hline
     $V_{1}^{D^{\ast}K^{\ast}}$ $=$ $0.74$
   & $V_{2}^{D^{\ast}K^{\ast}}$ $=$ $0.43$
   & $V_{3}^{D^{\ast}K^{\ast}}$ $=$ $0.26$
   & $V_{5}^{D^{\ast}K^{\ast}}$ $=$ $1.50$ \\
     $V_{6}^{D^{\ast}K^{\ast}}$ $=$ $0.78$
   & $A_{1}^{D^{\ast}K^{\ast}}$ $=$ $0.69$
   & $A_{3}^{D^{\ast}K^{\ast}}$ $=$ $0.15$
   & $A_{4}^{D^{\ast}K^{\ast}}$ $=$ $0.22$ \\ \hline
     $V^{D^{\ast}{\pi}}$     $=$ $0.92$
   & $A_{0}^{D^{\ast}{\pi}}$ $=$ $0.68$
   & $A_{1}^{D^{\ast}{\pi}}$ $=$ $0.74$
   & $A_{2}^{D^{\ast}{\pi}}$ $=$ $0.61$ \\ \hline
     $V^{D^{\ast}K}$         $=$ $1.04$
   & $A_{0}^{D^{\ast}K}$     $=$ $0.78$
   & $A_{1}^{D^{\ast}K}$     $=$ $0.85$
   & $A_{2}^{D^{\ast}K}$     $=$ $0.68$
   \end{tabular}
   \end{ruledtabular}
   \end{table}

   \section{Branching ratio}
   \label{sec:branch}
   The branching ratio of two-body nonleptonic $D^{\ast}$ decays
   is defined as
     \begin{equation}
    {\cal B}r\, =\,
     \frac{ G_{F}^{2} }{ 48\,{\pi} }\,
    {\vert}V_{cs}{\vert}^{2}\,{\vert}V_{ud}{\vert}^{2}\,
     \frac{p_{\rm cm}}{m_{D^{\ast}}^{2}\,{\Gamma}_{D^{\ast}}}\,
     \sum\limits_{s_{i}} {\cal M}{\cal M}^{\dagger}
    \label{eq:branch}.
    \end{equation}
   where $p_{\rm cm}$ is the center-of-mass momentum of
   final states in the rest frame of the $D^{\ast}$ meson,
   and ${\Gamma}_{D^{\ast}}$ is the decay width of
   the $D^{\ast}$ meson.
   ${\cal M}$ denotes the decay amplitude.
   For $D^{\ast}$ ${\to}$ $\overline{K}{\pi}^{+}$ decays,
   there is only the $p$-wave amplitude.
   For $D^{\ast}$ ${\to}$ $\overline{K}^{\ast}{\pi}^{+}$ and
   $\overline{K}{\rho}^{+}$ decays, there are $s$-, $p$-
   and $d$-wave amplitudes.
   The analytic expressions of decay amplitudes
   with NF approach are collected in
   Appendixes \ref{sec:v2pp} and \ref{sec:v2vp}.

   The decay width of the $D^{{\ast}+}$ meson has been experimentally
   determined to be ${\Gamma}_{D^{{\ast}+}}$ $=$
   $83.4{\pm}1.8$ keV \cite{pdg2020}, while there only was an
   upper limit for the decay width of the $D^{{\ast}0}$ meson,
   ${\Gamma}_{D^{{\ast}0}}$ $<$ $2.1$ MeV \cite{pdg2020}.
   It is widely assumed that there should be a relation between
   decay widths for the $p$-wave strong decay $D^{\ast}$ ${\to}$
   $D{\pi}^{0}$ \cite{PhysRevD.32.189,PhysRevD.32.810,PhysRevD.88.034034},
     \begin{equation}
     \frac{ {\Gamma}(D^{{\ast}0}{\to}D^{0}{\pi}^{0}) }
          { {\Gamma}(D^{{\ast}+}{\to}D^{+}{\pi}^{0}) }
     \, =\,
     \frac{ p_{{\rm cm}D^{0}{\pi}^{0}}^{3} }
          { p_{{\rm cm}D^{+}{\pi}^{0}}^{3} }\,
     \frac{ m_{D^{{\ast}+}}^{2} }
          { m_{D^{{\ast}0}}^{2} }
     \label{ratio-decay}.
     \end{equation}
   So it is expected that the decay width of $D^{{\ast}0}$,
     \begin{equation}
    {\Gamma}_{D^{{\ast}0}}
     \, =\,
    {\Gamma}_{D^{{\ast}+}}
     \frac{ {\cal B}r(D^{{\ast}+}{\to}D^{+}{\pi}^{0}) }
          { {\cal B}r(D^{{\ast}0}{\to}D^{0}{\pi}^{0}) }\,
     \frac{ p_{{\rm cm}D^{0}{\pi}^{0}}^{3} }
          { p_{{\rm cm}D^{+}{\pi}^{0}}^{3} }\,
     \frac{ m_{D^{{\ast}+}}^{2} }
          { m_{D^{{\ast}0}}^{2} }
     \, =\,
      55.9^{+5.9}_{-5.4}\,\text{keV}
     \label{decay-width-cuv},
     \end{equation}
   where the comprehensive uncertainties are conservative
   estimates, and come from data error of
   ${\Gamma}_{D^{{\ast}+}}$, branching ratios and mesonic mass.
   The full decay width ${\Gamma}_{D^{{\ast}0}}$ in
   Eq.(\ref{decay-width-cuv}) is well consistent with theoretical
   prediction of Ref. \cite{PhysRevD.88.034034}, and will be used
   in our calculation to estimate branching ratios for
   $D^{{\ast}0}$ decays.

   Our numerical calculation shows that branching ratios
   for nonleptonic $D^{\ast}$ decays are,
     \begin{eqnarray}
    {\cal B}r( D^{{\ast}+} {\to} \overline{K}^{0}{\pi}^{+} )
     & {\approx} & 1.6 {\times} 10^{-10}
     \label{br:d-pkz-pi}, \\
    {\cal B}r( D^{{\ast}+} {\to} \overline{K}^{{\ast}0}{\pi}^{+} )
     & {\approx} &  4.4 {\times} 10^{-10}
     \label{br:d-vkz-pi}, \\
    {\cal B}r( D^{{\ast}+} {\to} \overline{K}^{0}{\rho}^{+} )
     & {\approx} &  8.3 {\times} 10^{-10}
     \label{br:d-pkz-rho}, \\
    {\cal B}r( D^{{\ast}0} {\to} K^{-}{\pi}^{+} )
     & {\approx} &  7.3 {\times} 10^{-10}
     \label{br:u-pkm-pi}, \\
    {\cal B}r( D^{{\ast}0} {\to} K^{{\ast}-}{\pi}^{+} )
     & {\approx} & 2.0 {\times} 10^{-9}
     \label{br:u-vkm-pi}, \\
    {\cal B}r( D^{{\ast}0} {\to} K^{-}{\rho}^{+} )
     & {\approx} &  2.9 {\times} 10^{-9}
     \label{br:u-pkm-rho}.
     \end{eqnarray}

   Our comments on branching ratios are as follows.

   (1)
   The light valence quarks of the $D^{{\ast}0}$ and $D^{{\ast}+}$
   meson are usually call spectator quarks, and they do not take
   part in the weak interaction directly, when $W$ annihilation
   and $W$ exchange contributions are not considered with
   the NF approach.
   The partial width for $D^{{\ast}0}$ and $D^{{\ast}+}$ meson
   decays into the similar isospin final states should in principle
   be approximately equal.
   In fact, $D^{{\ast}+}$ meson decays are dynamically induced
   by both external and internal $W$ emission interactions, and
   the interference between these two contributions is
   destructive owing to a combination of Wilson coefficients.
   This is the main reason for he hierarchical relationship,
   ${\cal B}r( D^{{\ast}+} {\to} \overline{K}^{0}{\pi}^{+} )$
   $<$ ${\cal B}r( D^{{\ast}0} {\to} K^{-}{\pi}^{+} )$,
   ${\cal B}r( D^{{\ast}+} {\to} \overline{K}^{{\ast}0}{\pi}^{+} )$
   $<$ ${\cal B}r( D^{{\ast}0} {\to} K^{{\ast}-}{\pi}^{+} )$,
   and ${\cal B}r( D^{{\ast}+} {\to} \overline{K}^{0}{\rho}^{+} )$
   $<$ ${\cal B}r( D^{{\ast}0} {\to} K^{-}{\rho}^{+} )$.
   Similar hierarchical phenomena have also been observed
   experimentally in the Cabibbo-favored pseudoscalar $D^{0}$
   and $D^{+}$ meson decays, for example,
   ${\cal B}r( D^{+} {\to} K^{0}_{S}{\pi}^{+} )$
   $=$ $1.562(31)\%$ and
   ${\cal B}r( D^{0} {\to} K^{-}{\pi}^{+} )$ $=$
   $3.950(31)\%$ \cite{pdg2020}.

   (2)
   There are three partial wave amplitudes for $D^{\ast}$ ${\to}$
   $PV$ decays, while only the $p$-wave amplitude contributes to
   $D^{\ast}$ ${\to}$ $PP$ decays. Hence, there is a hierarchical
   relationship among branching ratios,
   ${\cal B}r( D^{{\ast}+} {\to} PP )$ $<$
   ${\cal B}r( D^{{\ast}+} {\to} PV )$ and
   ${\cal B}r( D^{{\ast}0} {\to} PP )$ $<$
   ${\cal B}r( D^{{\ast}0} {\to} PV )$.
   In addition, because of decay constants $f_{\rho}$ $>$ $f_{\pi}$,
   branching ratio for emission ${\rho}^{+}$ is generally
   larger than that for emission ${\pi}^{+}$,
   for either $D^{{\ast}0}$ or $D^{{\ast}+}$ decays.
   Similar phenomena have also been seen in pseudoscalar $D$
   meson decays, for example,
   ${\cal B}r( D^{0} {\to} K^{-}{\rho}^{+} )$
   $=$ $(11.3{\pm}0.77)\%$ and
   ${\cal B}r( D^{0} {\to} K^{{\ast}-}{\pi}^{+} )$
   $=$ $(2.31^{+0.40}_{-0.20})\%$
   determined from $D^{0}$ ${\to}$ $K^{-}{\pi}^{+}{\pi}^{0}$
   decays \cite{pdg2020}.

   (3)
   As it is well known the vector ${\rho}$ and
   $K^{\ast}$ mesons are resonances, and they will
   decay promptly into pseudoscalar mesons via the
   strong interactions, with branching ratios
   ${\cal B}r({\rho}{\to}{\pi}{\pi})$ ${\sim}$ $100\,\%$
   and ${\cal B}r(K^{\ast}{\to}K{\pi})$ ${\sim}$ $100\,\%$
   \cite{pdg2020}.
   The vector ${\rho}$ ($K^{\ast}$) meson in $D^{{\ast}0}$
   ${\to}$ $K^{-}{\rho}^{+}$ ($K^{{\ast}-}{\pi}^{+}$) decay
   should be reconstructed by the final pseudoscalar mesons.
   Besides, both $D^{{\ast}0}$ ${\to}$ $K^{-}{\rho}^{+}$
   and $D^{{\ast}0}$ ${\to}$ $K^{{\ast}-}{\pi}^{+}$ decays
   contribute to $D^{{\ast}0}$ ${\to}$ $K^{-}{\pi}^{+}{\pi}^{0}$.
   The branching ratio of the three-body
   $D^{{\ast}0}$ ${\to}$ $K^{-}{\pi}^{+}{\pi}^{0}$ decay can be
   approximately the sum of the branching ratios of these
   two two-body $D^{{\ast}0}$ ${\to}$ $PV$ modes whose
   interference effect happens
   in a small region of the Dalitz plot.
   A similar case can be seen by the comparison between the
   branching ratio
   ${\cal B}r(D^{0}{\to}K^{-}{\pi}^{+}{\pi}^{0})$ $=$
   $(14.4{\pm}0.5)\,\%$ and the sum of partial branching ratios
   ${\cal B}r(D^{0}{\to}K^{-}{\rho}^{+})$ $=$
   $(11.3{\pm}0.7)\,\%$ and
   ${\cal B}r(D^{0}{\to}K^{{\ast}-}{\pi}^{+})$ $=$
   $(2.31^{+0.40}_{-0.20})\,\%$ \cite{pdg2020}.
   Therefore, it may be an educated guess that the three-body
   $D^{{\ast}0}$ ${\to}$ $K^{-}{\pi}^{+}{\pi}^{0}$ decay will
   have a relatively larger branching ratio,
   ${\cal B}r(D^{{\ast}0}{\to}K^{-}{\pi}^{+}{\pi}^{0})$ ${\sim}$
   $5.0{\times}10^{-9}$, and can be more
   easily investigated in experiments, compared to the
   two-body $D^{{\ast}0}$ ${\to}$
   $K^{-}{\rho}^{+}$ and $K^{{\ast}-}{\pi}^{+}$ decays.

   (4)
   Theoretical predictions on branching ratios are easily influenced
   by a number of factors, including final state interactions and
   other contributions.
   For example, studies \cite{PhysRevD.67.014001,PhysRevD.81.074021}
   using the flavor topology diagrammatic approach have shown
   that for Cabibbo-favored $D$ meson decays, the $W$ exchange and
   annihilation contributions should deserve due attention,
   although the external $W$ emission contributions always give
   the largest contributions in the total amplitudes.
   We would like to point out that what we want is whether it is
   feasible to investigate the nonleptonic $D^{{\ast}}$ weak decays
   at the future experiments, so the magnitude order estimation
   rather than precise calculation on branching ratio may be enough.
   It is generally believed that the NF approach can give a reasonable
   and correct magnitude order estimation on branching ratio for
   nonleptonic heavy flavored meson decays arising from the
   external $W$ emission weak interactions.
   In this sense, the magnitude order of branching ratios in
   Eq.(\ref{br:d-pkz-pi}-\ref{br:u-pkm-rho}) seems to be
   reliable.
   The potential event numbers of the concerned $D^{\ast}$
   ${\to}$ $PP$, $PV$ decays are listed in
   Table \ref{tab:event-number}.
   It is clear that the Cabibbo-favored $D^{\ast}$ ${\to}$
   $\overline{K}{\pi}^{+}$, $\overline{K}^{\ast}{\pi}^{+}$,
   $\overline{K}{\rho}^{+}$ decays can be measurable
   at the future
   STCF, CEPC, FCC-ee and LHCb@HL-LHC experiments.
   The $D^{{\ast}+}$ ${\to}$ $K^{{\ast}-}{\pi}^{+}$ and
   $K^{-}{\rho}^{+}$ decays can also be investigated at
   SuperKEKB experiments.

  \begin{table}[th]
  \caption{The potential event numbers of $D^{\ast}$ ${\to}$
  $\overline{K}{\pi}^{+}$, $\overline{K}^{\ast}{\pi}^{+}$,
   $\overline{K}{\rho}^{+}$ decays at experiments, where the
  $D^{\ast}$ meson data available has been estimated in
  Section \ref{sec01}.}
  \label{tab:event-number}
  \begin{ruledtabular}
  \begin{tabular}{c|ccccc}
   experiment & SuperKEKB
              & STCF   & CEPC
              & FCC-ee & LHCb@HL-LHC \\ \hline
   $N_{D^{\ast}}$ & $2{\times}10^{10}$
              & $8{\times}10^{10}$    & $10^{11}$
              & $10^{12}$             & $2{\times}10^{14}$ \\ \hline
   $N_{ D^{{\ast}+} {\to} \overline{K}^{0}{\pi}^{+} }$
       & $3$ & $13$ & $16$ & $160$ & $3.2{\times}10^{4}$ \\
   $N_{ D^{{\ast}+} {\to} \overline{K}^{{\ast}0}{\pi}^{+} }$
       & $9$ & $35$ & $44$ & $440$ & $8.8{\times}10^{4}$ \\
   $N_{ D^{{\ast}+} {\to} \overline{K}^{0}{\rho}^{+} }$
       & $17$ & $66$ & $83$ & $830$ & $1.66{\times}10^{5}$ \\
   $N_{ D^{{\ast}0} {\to} K^{-}{\pi}^{+} }$
       & $14$ & $58$ & $72$ & $720$ & $1.44{\times}10^{5}$ \\
   $N_{ D^{{\ast}0} {\to} K^{{\ast}-}{\pi}^{+} }$
       & $40$ & $160$ & $200$ & $2000$ & $4.0{\times}10^{5}$ \\
   $N_{ D^{{\ast}0} {\to} K^{-}{\rho}^{+} }$
       & $57$ & $230$ & $287$ & $2870$ & $5.74{\times}10^{5}$ \\ 
   $N_{ D^{{\ast}0} {\to}K^{-}{\pi}^{+}{\pi}^{0} }$
       & $100$ & $400$ & $500$ & $5000$ & $1.0{\times}10^{6}$
   \end{tabular}
   \end{ruledtabular}
   \end{table}

  \section{Summary}
  \label{sec:summary}
   Now, more than 45 years after the discovery
   of the $D^{\ast}$ mesons, our knowledge and understanding
   of the nature of the $D^{\ast}$ mesons is far from enough,
   and needs to be substantially improved.
   One of the major reasons that excessively hindered
   experimental investigation on $D^{\ast}$ mesons is
   that data are too scarce.
   We should thank the high-luminosity particle physics experiments
   for offering us a huge amount of $D^{\ast}$ meson
   data and a tempting opportunity to explore the
   wanted $D^{\ast}$ meson in the future.
   Compared with the dominant $D^{\ast}$ ${\to}$ $D{\pi}$
   decays which are subject to kinematical factors,
   one advantage of nonleptonic $D^{\ast}$ weak decays
   is that the final pion and kaon mesons are energetic
   and easily detectable by the sensitive high-resolution
   detectors.
   In addition, study of nonleptonic $D^{\ast}$ weak decays
   is scientifically significant, and provide us with a
   new venue for testing SM.
   In this paper, the Cabibbo-favored two-body nonleptonic
   $D^{\ast}$ ${\to}$ $PP$, $PV$ decays were studied by
   using the NF approach within SM.
   It is found that branching ratios for $D^{\ast}$ ${\to}$
   $\overline{K}{\pi}^{+}$, $\overline{K}^{\ast}{\pi}^{+}$,
   $\overline{K}{\rho}^{+}$ decays can reach up to
   ${\cal O}(10^{-10})$ or more,
   and can be accessible at STCF, CEPC, FCC-ee and
   LHCb@HL-LHC experiments,
   which indicate that study of these weak interaction
   processes is experimentally feasible and practicable
   in the future.

  \section*{Acknowledgments}
  The work is supported by the National Natural Science Foundation
  of China (Grant Nos. 11705047, U1632109, 11875122) and Natural
  Science Foundation of Henan Province (Grant No. 222300420479),
  the Excellent Youth Foundation of Henan Province
  (Grant No. 212300410010).
  We would like to acknowledge the useful help and valuable
  discussion from Prof. Haibo Li (IHEP@CAS),
  Prof. Shuangshi Fang (IHEP@CAS),
  Prof. Frank Porter (Caltech),
  Prof. Antimo Palano (INFN),
  Prof. Chengping Shen (Fudan University),
  Dr. Ping Xiao (Fudan University),
  Dr. Qingping Ji (Henan Normal University),
  Dr. Huijing Li (Henan Normal University) and
  Ms. Liting Wang (Henan Normal University),
  and positive comments and constructive
  suggestions from referees.

  \begin{appendix}

   \section{amplitudes for $V$ ${\to}$ $P_{1}$ $+$ $P_{2}$ decays}
   \label{sec:v2pp}
   With the conventions of Eq.(\ref{decay-pseudoscalar-v}),
   Eq.(\ref{decay-pseudoscalar-a}),
   Eq.(\ref{fromfactor-v2p-v}) and
   Eq.(\ref{fromfactor-v2p-a}),
   the general expression of HMEs with NF approach for
   $V$ ${\to}$ $P_{1}$ $+$ $P_{2}$ transition
   can be written as,
     \begin{equation}
    {\cal M}
     \, =\,
    {\langle}P_{2}{\vert}V_{\mu}-A_{\mu}{\vert}0{\rangle}\,
    {\langle}P_{1}{\vert}V^{\mu}-A^{\mu}{\vert}V{\rangle}
     \, =\,
     {\cal M}_{p}\,({\epsilon}_{V}{\cdot}p_{P_{2}})
     \label{vpp-amp},
     \end{equation}
     \begin{equation}
    {\cal M}_{p}
     \, =\,
    2\, f_{P_{2}}\,m_{V}\,A_{0}^{VP_{1}}(0)
     \label{vpp-sub-amp-01},
     \end{equation}
     \begin{equation}
     \sum\limits_{s_{V}}
    {\cal M}\,{\cal M}^{\dagger}
     \, =\,
    {\vert}{\cal M}_{p}{\vert}^{2}\,p_{\rm cm}^{2}
     \label{vpp-amp-amp},
     \end{equation}
     \begin{equation}
     p_{\rm cm}
     \, =\,
    {\lambda}^{1/2}(m_{V}^{2},m_{P_{1}}^{2},m_{P_{2}}^{2})/2\,m_{V}
     \label{vpp-amp-pcm},
     \end{equation}
     \begin{equation}
    {\lambda}(a,b,c)
     \, =\,
    a^{2}+b^{2}+c^{2}-2\,a\,b-2\,b\,c-2\,c\,a
     \label{lambda-pcm}.
     \end{equation}

   With the NF approach,
   there is only one amplitude for the neutral $D^{{\ast}0}$
   meson decay in question, which corresponds to external $W$
   emission. There are two amplitudes for the charged
   $D^{{\ast}{\pm}}$ meson decay in question, which
   correspond to external and internal $W$ emissions.
   The partial wave amplitudes for $D^{{\ast}0}$ ${\to}$
   $K^{-}{\pi}^{+}$ decay are written as,
     \begin{equation}
    {\cal M}_{p}
     \, =\,
     2\, m_{D^{{\ast}}}\,
     a_{1}\, f_{\pi}\,A_{0}^{D^{{\ast}}K}
     \label{amp-cu-km-pip},
     \end{equation}
   and for $D^{{\ast}+}$ ${\to}$ $\overline{K}^{0}{\pi}^{+}$ decay,
     \begin{equation}
    {\cal M}_{p}
     \, =\,
     2\, m_{D^{{\ast}}}\, \big\{
     a_{1}\, f_{\pi}\,A_{0}^{D^{{\ast}}K}
    +a_{2}\, f_{K}\,A_{0}^{D^{{\ast}}{\pi}}
     \big\}
     \label{amp-cd-kz-pip},
     \end{equation}
   where the coefficients $a_{1}$ and $a_{2}$ correspond to
   external and internal $W$ emission, respectively; and
   they are defined as,
     \begin{equation}
    a_{1} \, =\, C_{1}+C_{2}/N_{c}
     \label{wc-a1},
     \end{equation}
     \begin{equation}
    a_{2} \, =\, C_{2}+C_{1}/N_{c}
     \label{wc-a2}.
     \end{equation}
   In practice, it is generally believed that coefficients $a_{1,2}$
   are also influenced by nonfactorizable contributions and
   final state interactions.
   In many phenomenological studies on charmed meson weak decays,
   such as  Refs. \cite{zpc.34.103,
   Stech.1985,npb.133.315,cpc.27.665,epjc.42.391,
   PhysRevD.84.074019,PhysRevD.81.074021,PhysRevD.93.114010,
   PhysRevD.86.036012}, coefficients
   $a_{1}$ ${\approx}$ $1.2$ and $a_{2}$ ${\approx}$ $-0.5$
   are often used for charmed meson decays by including
   comprehensive contributions.

   \section{amplitudes for $V_{1}$ ${\to}$ $V_{2}$ $+$ $P$ decays}
   \label{sec:v2vp}
   With the conventions of Eqs.(\ref{decay-pseudoscalar-v}-\ref{fromfactor-v2v-a}),
   the general expression of HMEs with NF approach for
   $V_{1}$ ${\to}$ $V_{2}$ $+$ $P$ transition
   can be written as,
     \begin{eqnarray}
    {\cal M} &=&
    {\langle}V_{2}{\vert}V_{\mu}-A_{\mu}{\vert}0{\rangle}\,
    {\langle}P    {\vert}V_{\mu}-A_{\mu}{\vert}V_{1}{\rangle}
     \nonumber \\ &=&
    {\cal M}_{s}\,
     ( {\epsilon}_{V_{1}}{\cdot}{\epsilon}_{V_{2}}^{\ast} )
   + \frac{ {\cal M}_{d} }{ m_{V_{1}}\,m_{V_{2}} }\,
     ( {\epsilon}_{V_{1}}{\cdot}p_{V_{2}} ) \,
     ( {\epsilon}_{V_{2}}^{\ast}{\cdot}p_{V_{1}} )
     \nonumber \\ & +&
     \frac{ {\cal M}_{p} }{ m_{V_{1}}\,m_{V_{2}} }\,
     {\varepsilon}_{{\mu}{\nu}{\alpha}{\beta}}\,
     {\epsilon}_{V_{1}}^{\mu}\,
     {\epsilon}_{V_{2}}^{{\ast}{\nu}}\,
      p_{V_{1}}^{\alpha}\, p_{V_{2}}^{\beta}
     \label{vpv-amp},
     \end{eqnarray}
     \begin{eqnarray}
    {\cal M}^{\prime} &=&
    {\langle}P    {\vert}V_{\mu}-A_{\mu}{\vert}0{\rangle}\,
    {\langle}V_{2}{\vert}V_{\mu}-A_{\mu}{\vert}V_{1}{\rangle}
     \nonumber \\ &=&
    {\cal M}_{s}^{\prime}\,
     ( {\epsilon}_{V_{1}}{\cdot}{\epsilon}_{V_{2}}^{\ast} )
   + \frac{ {\cal M}_{d}^{\prime} }{ m_{V_{1}}\,m_{V_{2}} }\,
     ( {\epsilon}_{V_{1}}{\cdot}p_{V_{2}} ) \,
     ( {\epsilon}_{V_{2}}^{\ast}{\cdot}p_{V_{1}} )
     \nonumber \\ & +&
     \frac{ {\cal M}_{p}^{\prime} }{ m_{V_{1}}\,m_{V_{2}} }\,
     {\varepsilon}_{{\mu}{\nu}{\alpha}{\beta}}\,
     {\epsilon}_{V_{1}}^{\mu}\,
     {\epsilon}_{V_{2}}^{{\ast}{\nu}}\,
      p_{V_{1}}^{\alpha}\, p_{V_{2}}^{\beta}
     \label{vvp-amp},
     \end{eqnarray}
     \begin{eqnarray}
    {\cal M}_{s} &=&
     -i\, f_{V_{2}}\,m_{V_{2}}\,( m_{V_{1}}+m_{P} )\
        A_{1}^{V_{1}P}(0)
     \label{vpv-amp-s}, \\
    {\cal M}_{d} &=&
    -i\,f_{V_{2}}\,m_{V_{1}}\,m_{V_{2}} \,
      \frac{ 2\,m_{V_{2}}  }
           { m_{V_{1}}+m_{P} }\, A_{2}^{V_{1}P}(0)
     \label{vpv-amp-d}, \\
     {\cal M}_{p} &=&
     - f_{V_{2}}\,m_{V_{1}}\,m_{V_{2}}  \,
             \frac{ 2\,m_{V_{2}} }
                  { m_{V_{1}}+m_{P} }\, V^{V_{1}P}(0)
     \label{vpv-amp-p},
     \end{eqnarray}
     \begin{eqnarray}
    {\cal M}_{s}^{\prime} &=&
     -i\, f_{P}\,(m_{V_{1}}^{2}-m_{V_{2}}^{2})\,
      V_{1}^{V_{1}V_{2}}(0)
     \label{vvp-amp-s}, \\
    {\cal M}_{d}^{\prime} &=&
    -i\,f_{P}\, m_{V_{1}}\,m_{V_{2}}\,
             \big\{ V_{4}^{V_{1}V_{2}}(0)
                   -V_{5}^{V_{1}V_{2}}(0)
                   +V_{6}^{V_{1}V_{2}}(0) \big\}
     \label{vvp-amp-d}, \\
     {\cal M}_{p}^{\prime} &=&
    -2\,f_{P}\,m_{V_{1}}\,m_{V_{2}}\,A_{1}^{V_{1}V_{2}}(0)
     \label{vvp-amp-p},
     \end{eqnarray}
     \begin{eqnarray}
     \sum\limits_{s_{V_{1}}}
     \sum\limits_{s_{V_{2}}}
    {\cal M}\,{\cal M}^{\dagger} &=&
    {\vert}{\cal M}_{s}{\vert}^{2}\,(x^{2}+2)
   +2\,{\cal R} \big({\cal M}_{s}\,
       {\cal M}_{d}^{\ast}\big)
        \,x\,(x^{2}-1)
     \nonumber \\ & +&
    {\vert}{\cal M}_{d}{\vert}^{2}\,(x^{2}-1)^{2}
   +2\,{\vert}{\cal M}_{p}{\vert}^{2}\,(x^{2}-1)
     \label{vpv-amp-amp},
     \end{eqnarray}
     \begin{eqnarray}
     \sum\limits_{s_{V_{1}}}
     \sum\limits_{s_{V_{2}}}
    {\cal M}^{\prime}\,{\cal M}^{{\prime}\dagger} &=&
    {\vert}{\cal M}_{s}^{\prime}{\vert}^{2}\,(x^{2}+2)
   +2\,{\cal R} \big({\cal M}_{s}^{\prime}\,
       {\cal M}_{d}^{{\prime}{\ast}}\big)
        \,x\,(x^{2}-1)
     \nonumber \\ & +&
    {\vert}{\cal M}_{d}^{\prime}{\vert}^{2}\,(x^{2}-1)^{2}
   +2\,{\vert}{\cal M}_{p}^{\prime}{\vert}^{2}\,(x^{2}-1)
     \label{vvp-amp-amp},
     \end{eqnarray}
     \begin{equation}
     x \, =\, \frac{ p_{V_{1}}{\cdot}p_{V_{2}} }
                   { m_{V_{1}}\,m_{V_{2}} }
       \, =\, \frac{ m_{V_{1}}^{2}+m_{V_{2}}^{2}-m_{P}^{2} }
                   { 2\,m_{V_{1}}\,m_{V_{2}} }
     \label{vpv-amp-x}.
     \end{equation}

   For $D^{{\ast}0}$ ${\to}$ $K^{{\ast}-}{\pi}^{+}$ decay,
   the partial wave amplitudes are written as,
     \begin{eqnarray}
    {\cal M}_{s}^{\prime} &=&
     -i\, f_{\pi}\, a_{1}\, (m_{D^{\ast}}^{2}-m_{K^{\ast}}^{2})\,
       V_{1}^{D^{\ast}K^{\ast}}
     \label{amp-cu-kvm-pip-s}, \\
    {\cal M}_{d}^{\prime} &=&
    -i\,f_{\pi}\, a_{1}\, m_{D^{\ast}}\,m_{K^{\ast}}\,
             \big\{ V_{4}^{D^{\ast}K^{\ast}}
                   -V_{5}^{D^{\ast}K^{\ast}}
                   +V_{6}^{D^{\ast}K^{\ast}} \big\}
     \label{amp-cu-kvm-pip-d}, \\
     {\cal M}_{p}^{\prime} &=&
    -2\,f_{\pi}\,a_{1}\, m_{D^{\ast}}\,m_{K^{\ast}}\,
       A_{1}^{D^{\ast}K^{\ast}}
     \label{amp-cu-kvm-pip-p}.
     \end{eqnarray}

   For $D^{{\ast}0}$ ${\to}$ $K^{-}{\rho}^{+}$ decay,
   the partial wave amplitudes are written as,
     \begin{eqnarray}
    {\cal M}_{s} &=&
     -i\, f_{\rho}\, a_{1}\,
       m_{\rho}\,( m_{D^{\ast}}+m_{K} )\
       A_{1}^{D^{\ast}K}
     \label{amp-cu-km-rhop-s}, \\
    {\cal M}_{d} &=&
    -i\, f_{\rho}\, a_{1}\, m_{D^{\ast}} \, m_{\rho}\,
      \frac{ 2\,m_{\rho}\,  }
           {  m_{D^{\ast}}+m_{K} }\, A_{2}^{D^{\ast}K}
     \label{amp-cu-km-rhop-d}, \\
     {\cal M}_{p} &=&
    - f_{\rho}\, a_{1}\, m_{D^{\ast}} \, m_{\rho}\,
      \frac{ 2\,m_{\rho}\,  }
           {  m_{D^{\ast}}+m_{K} }\, V^{D^{\ast}K}
     \label{amp-cu-km-rhop-p}.
     \end{eqnarray}

   For $D^{{\ast}+}$ ${\to}$ $\overline{K}^{{\ast}0}{\pi}^{+}$
   decay, each of partial wave amplitudes can be divided into
   two parts,
     \begin{equation}
    {\cal M}_{i}\, =\, {\cal M}_{i}^{(1)}+{\cal M}_{i}^{(2)},
     \quad \text{for}\ i\,=\,s,p,d.
     \label{amp-cd-kvz-pip-two-parts}
     \end{equation}
     \begin{eqnarray}
    {\cal M}_{s}^{(2)} &=&
     -i\, f_{K^{\ast}}\, a_{2}\,
       m_{K^{\ast}}\,( m_{D^{\ast}}+m_{\pi} )\
       A_{1}^{D^{\ast}{\pi}}
     \label{amp-cd-kvz-pip-s}, \\
    {\cal M}_{d}^{(2)} &=&
    -i\, f_{K^{\ast}}\, a_{2}\, m_{D^{\ast}} \, m_{K^{\ast}}\,
      \frac{ 2\,m_{K^{\ast}}\,  }
           {  m_{D^{\ast}}+m_{\pi} }\,
       A_{2}^{D^{\ast}{\pi}}
     \label{amp-cd-kvz-pip-d}, \\
     {\cal M}_{p}^{(2)} &=&
    - f_{K^{\ast}}\, a_{2}\, m_{D^{\ast}} \, m_{K^{\ast}}\,
      \frac{ 2\,m_{K^{\ast}}  }
           {  m_{D^{\ast}}+m_{\pi} }\,
       V^{D^{\ast}{\pi}}
     \label{amp-cd-kvz-pip-p},
     \end{eqnarray}
   and expressions of ${\cal M}_{i}^{(1)}$ are the same
   as those of ${\cal M}_{s,p,d}^{\prime}$ for $D^{{\ast}0}$
   ${\to}$ $K^{{\ast}-}{\pi}^{+}$ decay in
   Eq.(\ref{amp-cu-kvm-pip-s}),
   Eq.(\ref{amp-cu-kvm-pip-d}) and
   Eq.(\ref{amp-cu-kvm-pip-p}).

   For $D^{{\ast}+}$ ${\to}$ $\overline{K}^{0}{\rho}^{+}$
   decay, each of partial wave amplitudes can also be divided into
   two parts similar to Eq.(\ref{amp-cd-kvz-pip-two-parts}),
     \begin{eqnarray}
    {\cal M}_{s}^{(2)} &=&
     -i\, f_{K}\, a_{2}\, (m_{D^{\ast}}^{2}-m_{\rho}^{2})\,
       V_{1}^{D^{\ast}{\rho}}
     \label{amp-cd-kz-rhop-sp}, \\
    {\cal M}_{d}^{(2)} &=&
    -i\,f_{K}\, a_{2}\, m_{D^{\ast}}\,m_{\rho}\,
             \big\{ V_{4}^{D^{\ast}{\rho}}
                   -V_{5}^{D^{\ast}{\rho}}
                   +V_{6}^{D^{\ast}{\rho}} \big\}
     \label{amp-cd-kz-rhop-dp}, \\
     {\cal M}_{p}^{(2)} &=&
    -2\,f_{K}\,a_{2}\, m_{D^{\ast}}\,m_{\rho}\,
       A_{1}^{D^{\ast}{\rho}}
     \label{amp-cd-kz-rhop-pp},
     \end{eqnarray}
   and expressions of ${\cal M}_{i}^{(1)}$ are the same
   as those of ${\cal M}_{s,p,d}$ for $D^{{\ast}0}$
   ${\to}$ $K^{-}{\rho}^{+}$ decay in
   Eq.(\ref{amp-cu-km-rhop-s}),
   Eq.(\ref{amp-cu-km-rhop-d}) and
   Eq.(\ref{amp-cu-km-rhop-p}).

  \end{appendix}


  \end{document}